\documentclass[twocolumn,aps,prd,amsmath,amssymb]{revtex4}
\bibpunct{}{}{,}{s}{}{,}

\newtheorem{theorem}{Theorem}[section]
\newtheorem{corollary}[theorem]{Corollary}

\renewcommand\l{\lambda}
\newcommand{\A}{{\hat A}}
\renewcommand{\P}{{\hat P}}
\newcommand{\Hi}{{\cal H}}
\newcommand\BH{\mathcal{B(H)}}

\newcommand\PV{\mathcal{P}(V)}

\newcommand\dastoo[2]{\delta^o(\hat{#2})_{#1}}  
\newcommand\dasB[1]{\breve{\delta}(\hat{#1})} 
\newcommand{\SR}{\ps{\mathbb{R}^{\succeq}}} 	

\newcommand\ps[1]{\underline{#1}}        
\newcommand\Sig{\ps{\Sigma}}

\newcommand\Sub[1]{{\rm Sub}(#1)}              
\newcommand\Subcl[1]{{\rm Sub}_{{\rm cl}}(#1)} 

\newcommand\Set{{\bf Sets}}                   
\newcommand\SetC[1]{\Set^{{#1}^{\rm op}}}     
\newcommand\SetVNop{\SetC{\mathcal{V(N)}}}		

\newcommand\VN{\mathcal{V(N)}}
\newcommand\N{\mathcal{N}}
\newcommand\PN{\mathcal{P(N)}}
\newcommand\wpsi{\ps{\mathfrak{w}}^\psi}
\newcommand\Rlr{\ps{\mathbb{R}^\leftrightarrow}}

\begin{document}

\title{Quantum States and Measures on the Spectral Presheaf}

\author{Andreas D\"{o}ring}
\email{a.doering@imperial.ac.uk}
\affiliation{Theoretical Physics Group, Blackett Laboratory, Imperial College London}

\begin{abstract}
After a brief introduction to the spectral presheaf, which serves as an analogue of state space in the topos approach to quantum theory, we show that every state $\rho$ of the von Neumann algebra $\N$ of physical quantities of a quantum system determines a certain measure $\mu_\rho$ on the spectral presheaf of the system. The so-called clopen subobjects of the spectral presheaf play the role of measurable sets. Measures on the spectral presheaf can be characterised abstractly, and the main result is that every abstract measure $\mu$ induces a unique state $\rho_\mu$ of the von Neumann algebra $\N$. Finally, we show how quantum-theoretical expectation values can be calculated from measures associated to quantum states.

\bigskip

\end{abstract}

\maketitle

\section{Introduction}
In the search for a theory of quantum gravity, we will have to consider radically new concepts, both physically and mathematically. Many researchers hope that conceptual breakthroughs will be found `on the way', while the starting points for the development of a theory of quantum gravity -- quantum theory and general relativity -- remain largely unchanged. (On the whole, there seems to be more willingness to try to adapt general relativity to quantum theory than the other way around.) Despite huge efforts, these hopes have never materialised. The well-known conceptual and interpretational problems of quantum theory, some of which become more severe in combination with general relativity, may necessitate a massive revision or reformulation of quantum theory itself before progress in the direction of quantum gravity can be made.

Unfortunately, the usual Hilbert space formalism does not allow for any obvious changes which would provide different and/or more general theories. It is hardly possible to give up on some part of the mathematical structure of the Hilbert space formalism without destroying the whole edifice. It may be necessary to find a radically new mathematical framework for quantum theory and other theories of physics beyond that.

The topos approach as initiated by Isham and Butterfield \cite{IB98,IB99,IB00,IB01,IB00b,IB02} and developed by Isham and the author \cite{DI(1),DI(2),DI(3),DI(4),Doe07b,DI08} is a proposal for such a new framework for physical theories. In the papers mentioned above, it was shown that large parts of quantum theory can be expressed using mathematical structures within a suitable topos of presheaves. These results go beyond a mere reformulation of quantum theory and provide a wealth of new structures to be explored. Moreover, the topos approach allows for generalisations beyond quantum theory. For all the details on this approach, see ref. \cite{DI08}.

Currently, many interesting open questions remain, even with respect to standard quantum theory. In this article, we will concentrate on the \emph{spectral presheaf}, which is a mathematical object that takes the role of the state space of a quantum system in the topos approach. Here, state space is to be seen in analogy to classical physics and does not mean Hilbert space. In fact, the spectral presheaf is not a space or set at all, as will be explained in section \ref{SecSpecPresh}. We will then show in section \ref{SecProposAsSubobjs} how propositions about the values of physical quantities correspond to subobjects of the spectral presheaf. Section \ref{SecMeasures} shows how states of a quantum system determine probability measures on the spectral presheaf. This is in analogy to classical physics, where states \emph{are} probability measures on state space. Measures are then characterised abstractly, and it is shown that from every abstract measure $\mu$, a unique state $\rho_\mu$ of the quantum system can be reconstructed. Thus there is a bijection between the space $\mathcal{S}(\N)$ of states of the quantum system and the set $\mathcal{M}(\Subcl{\Sig})$ of measures on the (clopen subobjects of the) spectral presheaf. The expectation values of ordinary quantum theory can be expressed in terms of measures, as is shown in section \ref{SecMeasExpVals}. In this section and in the discussion (section \ref{SecDiscussion}), we point out a number of open questions. 

While the mathematical structures that we consider are lying in a topos of presheaves, neither topos theory itself nor any serious category theory is needed in this article. For the experts, it should be mentioned that our arguments are topos-external. It will be useful to develop many aspects internally, but this is a task for the future. What we need here are some basics of the theory of von Neumann algebras and little more than the definitions of a category and a presheaf. Wherever possible, physical interpretations of the mathematical structures are given, and the analogy between the spectral presheaf of a quantum system and the state space of a classical system is emphasised. We hope this article can serve as an introduction for physicists to some of the structures provided by the topos approach to quantum theory.


\section{The spectral presheaf}			\label{SecSpecPresh}
The spectral presheaf was originally defined by Isham and Butterfield \cite{IB98,IB99,IB00,IB02} and was used extensively by Isham and the author in the development of a topos version of quantum theory \cite{DI(1),DI(2),DI(3),DI(4),Doe07b,DI08}. We will only give a short introduction here. For details, we refer the reader to the references mentioned above.

\subsection{Von Neumann algebras and contexts}
Consider a quantum system $S$. A basic assumption is that the physical quantities of $S$ can be described by the self-adjoint operators in a non-commutative \emph{von Neumann algebra} $\N$. For the theory of von Neumann algebras, see refs.\cite{KR83,Tak79+02}. An important example of a von Neumann algebra is the algebra $\BH$ of all bounded operators on a Hilbert space $\Hi$. We will only consider separable Hilbert spaces. It is well-known that every von Neumann algebra $\N$ is given as a subalgebra of some $\BH$ for a suitable Hilbert space $\Hi$. If $\mathcal{A}$ is some set of operators, then the \emph{commutant} of $\mathcal{A}$ in $\BH$ is defined as
\begin{equation}
			\mathcal{A}':=\{\hat B\in\BH\mid\forall\hat A\in\mathcal{A}:[\hat B,\hat A]=\hat 0\}.
\end{equation}
Let $\mathcal{A}$ be a set of operators which is closed under taking adjoints and which contains the identity operator $\hat 1$. Then the double commutant
\begin{equation}
			\mathcal{A}''
\end{equation}
is a von Neumann algebra. (A von Neumann algebra always contains the identity operator $\hat 1$ on $\Hi$.) We sometimes apply the double commutant construction to a set of projection operators, e.g. $\{\P,\hat 1\}$. The double commutant construction amounts to taking all Borel functions of the operators in the set. In particular, the difference $\hat 1-\P$ and the null projection $\hat 0$ are contained in the double commutant $\{\P,\hat 1\}''$.

If the set $\mathcal{A}$ contains only commuting operators (like, for example, $\{\P,\hat 1\}$ does), then $\mathcal{A}''$ is a commutative von Neumann algebra. The self-adjoint operators in $\mathcal{A}''$ represent commuting physical quantities. A commutative von Neumann subalgebra of the non-commutative von Neumann algebra $\N$ describing our quantum system is called a \emph{context}. We assume that every context is unital, i.e., contains the identity operator $\hat 1$. Contexts are typically denoted as $V,V_1,V_2,...$.

A context can be understood as a `classical snapshot', or a `classical perspective' on the quantum system $S$. Of course, no single context $V$ can give a complete picture of the quantum system. It is part of the conceptual scheme of the topos approach to quantum theory to consider \emph{all} contexts at the same time and to treat them in a uniform manner. In this way, one may hope to obtain a complete picture of the quantum system. 

The set of contexts, i.e., commutative unital von Neumann subalgebras of $\N$, is denoted as $\VN$. We exclude the trivial context $V_0=\mathbb{C}\hat 1$ from $\VN$. The set of contexts becomes partially ordered under inclusion of smaller contexts into larger ones. Typically, we denote the larger context as $V$ and the smaller one as $V'$ (which in this case does \emph{not} mean the commutant of $V$). As a partially ordered set, $\VN$ also is a category \cite{McL98} in a simple way: the objects of $\VN$ are the contexts $V$, and the arrows are the inclusions, e.g. $i_{V'V}:V'\rightarrow V$. We call $\VN$ the \emph{context category}.

Going from a larger context $V$ to a smaller context $V'$ can be understood as a process of \emph{coarse-graining}: the smaller algebra $V'$ contains less self-adjoint operators than $V$, so one can describe less physics from the perspective of $V'$. In this sense, $V'$ is coarser than $V$. The context category encodes this information, but it does more: let $V_1,V_2$ be two different contexts such that their intersection $V':=V_1\cap V_2$ is not just the trivial commutative algebra $\mathbb{C}\hat 1$. Then there are inclusion arrows from $V'$ to $V_1$ and from $V'$ to $V_2$. In this indirect way, the contexts $V_1$ and $V_2$ are related. While this structure may seem quite weak, it is behind powerful theorems like Gleason's theorem \cite{Gle57}, its generalisation to von Neumann algebras \cite{Mae89}, the Kochen-Specker theorem \cite{KS67} and its generalisation to von Neumann algebras.\cite{Doe05} In these theorems, conditions are posed only on commuting projections resp. commuting self-adjoint operators, yet one obtains results pertaining to the whole non-commutative von Neumann algebra of physical quantities. The reason is that each projection resp. self-adjoint operator typically is contained in many contexts, which are related in the way described for $V_1$ and $V_2$ above. Thus, one actually obtains conditions across contexts, which ultimately allow to prove the theorems. (In this article, we will use the generalised version of Gleason's theorem in section \ref{SecMeasures}.)

We will often refer to a context $V$ (or some structure associated to it) as `local', while the whole, non-commutative von Neumann algebra $\N$ is the `global' structure. Local and global thus do not refer to space or space-time in this article.  


\subsection{Gel'fand spectra and the spectral presheaf}
Each commutative von Neumann algebra is a commutative $C^*$-algebra and, as such, has a \emph{Gel'fand spectrum}. Let $V$ be a context, then the Gel'fand spectrum $\Sigma_V$ of $V$ is the set of algebra homomorphisms $\l:V\rightarrow\mathbb{C}$. Each $\l\in\Sigma_V$ can also be seen as a positive linear functional on $V$ of norm $1$ that is multiplicative, i.e., for all $\A,\hat B$, one has
\begin{eqnarray}
			\l(\A\hat B)=\l(\A)\l(\hat B).			
\end{eqnarray}
For each projection $\P$ in the context $V$, we get
\begin{equation}
			\l(\P)=\l(\P^2)=\l(\P)^2\in\{0,1\}.
\end{equation}
If the projections in $V$ are interpreted as representing \emph{propositions} of the form ``$A\epsilon\Delta$'', that is, ``the physical quantity $A$ (which is represented by a self-adjoint operator $\A\in V$) has a value in the (Borel) set $\Delta$'', then each element $\l$ of the Gel'fand spectrum $\Sigma_V$ can be seen as an assignment of truth-values to these propositions. If we have $\l(\P)=1$ for a projection, then the corresponding proposition ``$A\epsilon\Delta$'' is true, and if $\l(\P)=0$, then the proposition is false. (To be precise, to each projection $\P$, there correspond many propositions ``$A\epsilon\Delta$'', ``$B\epsilon\Gamma$'', ...) The existence of true-false truth-value assigments is peculiar to the commutative situation (with one exception): the Kochen-Specker theorem \cite{KS67} and its generalisation \cite{Doe05} show that such truth-value assignments are impossible for the projections in a non-commutative von Neumann algebra $\N$, with the exception of type $I_2$-algebras only. (The latter are $2\times2$-matrix algebras with entries from a commutative algebra. For such algebras, truth-value assignments exist.)
  
The Gel'fand spectrum $\Sigma_V$, equipped with the weak* topology, is a compact Hausdorff space. Since $V$ is a von Neumann algebra, $\Sigma_V$ is extremely disconnected. The \emph{clopen} (that is, closed and open) subsets of $\Sigma_V$ form a base of the topology (see thm. 3.2 in ref. \cite{deG05c}). There exists a lattice isomorphism
\begin{equation}			\label{IsomP(V)Subcl(Sigma_V)}
			\alpha:\mathcal{P}(V)\longrightarrow\Subcl{\Sigma_V}
\end{equation}
between the lattice of projections in $V$ and the lattice of clopen subsets of the Gel'fand spectrum $\Sigma_V$ of $V$. It is given, for all $\P\in\mathcal{P}(V)$, by
\begin{equation}
			\alpha(\P)=\{\l\in\Sigma_V\mid\l(\P)=1\}.
\end{equation}
Both $\mathcal{P}(V)$ and $\Subcl{\Sigma_V}$ are complete, distributive, orthocomplemented (i.e., Boolean) lattices.

As a $C^*$-algebra, $V$ is isomorphic to the $C^*$-algebra $C(\Sigma_V)$ of continuous, complex-valued functions on $\Sigma_V$. This isomorphism sends each operator $\A\in V$ to its \emph{Gel'fand transform} $\overline{A}$, given for all $\l\in\Sigma_V$ by
\begin{equation}
			\overline{A}(\l):=\l(\A).
\end{equation}
A self-adjoint operator $\A$ is sent to a real-valued function $\overline{A}$. Moreover, $\l(\A)\in\operatorname{sp}(\A)$, so each element $\l$ of the Gel'fand spectrum $\Sigma_V$ assigns values to all physical quantities in the context $V$.

The physical interpretation is the following: the Gel'fand spectrum $\Sigma_V$ can be seen as a `state space locally at $V$', and the physical quantities in $V$ (and only those!) are represented by real-valued functions on $\Sigma_V$. This is very close to classical physics, where one also has a state space, and physical quantities are represented by real-valued functions on it.

If $V'\subset V$ is a smaller context, then there is a canonical way to define a function from $\Sigma_V$ to $\Sigma_{V'}$, simply by restriction:
\begin{eqnarray}			\label{DefSig(i_VPrimeV)}
			\Sigma(i_{V'V}):\Sigma_V &\longrightarrow& \Sigma_{V'}\\			\nonumber
			\l &\longmapsto& \l|_{V'}.
\end{eqnarray}
This function is continuous, closed and open (see thm. 15 in ref. \cite{DI08}). In this way, the local state spaces belonging to different contexts are related in a non-trivial way.

The \emph{spectral presheaf} $\Sig$ is nothing but the collection of all these local state spaces $\Sigma_V$ for all $V\in\VN$, together with the functions $\Sigma(i_{V'V})$ between them. Technically, the spectral presheaf is a contravariant functor from the context category $\VN$ to the category $\Set$ of sets and functions. Such a functor is called a \emph{presheaf}.\cite{MM92} In our notation, presheaves will always be underlined. Concretely, the spectral presheaf is given
\begin{enumerate}
	\item on objects: for all $V\in\VN$, the component $\Sig_V$ is $\Sigma_V$, the Gel'fand spectrum of $V$;
	\item on arrows: for all inclusions $i_{V'V}:V'\rightarrow V$, the function $\Sig(i_{V'V})$ is $\Sigma(i_{V'V})$ as defined in (\ref{DefSig(i_VPrimeV)}).
\end{enumerate}
The spectral presheaf $\Sig$ takes the role of the state object for quantum theory in the topos approach \cite{IB98, IB00,DI(2),DI(3),DI(4),DI08}. It is interpreted as an analogue of state space. Being a presheaf, $\Sig$ is \emph{not} a space or set, but rather a whole collection of sets, more precisely, compact Hausdorff spaces $\Sig_V$, together with continuous functions between these spaces whenever a smaller context $V'$ is contained in a larger context $V$. In this manner, the definition of $\Sig$ takes contextuality into account.

Clearly, the spectral presheaf $\Sig$ depends on the quantum system, because $\Sig$ is constructed from the algebra $\N$ of physical quantities of the system. The algebra of physical quantities is a much finer invariant of a quantum system than Hilbert space. (As is well-known, all separable Hilbert spaces are isomorphic.) The relation between $\Sig$ and Hilbert space is rather indirect: the von Neumann algebra $\N$ is the subalgebra of some $\BH$, but this does not play an important role in the construction. On the other hand, it is of course useful to think of the elements of $\N$ as operators on some Hilbert space, and we are constantly making use of this, in particular with respect to projections. 

The spectral presheaf itself is one object in a category whose objects are presheaves, namely presheaves over the context category $\VN$. The category $\SetVNop$ of presheaves over $\VN$, with natural transformations between presheaves as arrows, is a \emph{topos}. Among other things, this means that $\SetVNop$ has an \emph{internal logic}. This logic is non-Boolean, it is of intuitionistic type, which means that the law of excluded middle does not hold. An important conceptual aspect of the topos approach is the use of this internal logic in defining a new form of quantum logic. As a consequence, topos quantum logic is intuitionistic. Very importantly, one has a distributive logic with a powerful deductive system given by the topos, which is very different from ordinary Birkhoff-von Neumann quantum logic. \cite{BvN36}


\section{Propositions as subobjects}			\label{SecProposAsSubobjs}
We will present the most direct approach to defining representatives of propositions in a topos form of quantum logic. More details can be found in ref. \cite{DI(2)}. The propositions we are concerned with are of the form ``$A\epsilon\Delta$'', i.e., ``the physical quantity $A$ has a value in the set $\Delta$''. Here, $A$ is some physical quantity of the quantum system under consideration, and $\Delta$ is some Borel subset of the real numbers. As usual, in quantum theory a physical quantity $A$ is represented by a self-adjoint operator $\A$ in the von Neumann algebra of physical quantities of the system, and the spectral theorem shows that a proposition ``$A\epsilon\Delta$'' is represented by a projection operator $\P=\hat E[A\epsilon\Delta]$.

In classical physics, physical quantities are represented by real-valued functions (which are measurable at least) on the state space of the system, and propositions ``$A\epsilon\Delta$'' are represented by (measurable) subsets of state space. Analogously, one may expect that in the topos form of quantum theory, propositions are represented by subobjects of the state object $\Sig$.

As a first step, consider some projection $\P\in\PN$, which represents a suitable proposition ``$A\epsilon\Delta$''. Some contexts $V\in\VN$ contain $\P$, but most of them do not. In every context $V\in\VN$, we approximate $\P$ from above by the smallest projection contained in $V$ larger than or equal to $\P$: 
\begin{equation}			\label{DefOuterDas}
			\dastoo{V}{P}:=\bigwedge\{\hat Q\in\mathcal{P}(V)\mid\hat Q\geq\P\}.
\end{equation}
If $\P\in\PV$, then $\dastoo{V}{P}=\P$, otherwise, $\dastoo{V}{P}>\P$. This is one instance of the concept of coarse-graining: the proposition/projection $\P$ is adapted to every context $V$. From a single projection $\P$, we obtain a whole collection of projections $\dastoo{V}{P}$, one for each context $V\in\VN$.

We now define a mapping from the lattice $\PN$ of projections in $\N$ to the subobjects of $\Sig$, called \emph{daseinisation of projections}:
\begin{eqnarray}			\label{DefDasOfProjs}
			\ps{\delta}:\PN &\longrightarrow& \Sub{\Sig}\\			\nonumber
			\P &\longmapsto& (\alpha(\dastoo{V}{P}))_{V\in\VN}=:\ps{\delta(\P)},
\end{eqnarray}
where, for each $V\in\VN$, we use the isomorphism (\ref{IsomP(V)Subcl(Sigma_V)}) between the projections in $V$ and the clopen subsets of $\Sig_V$. We thus get a collection of (clopen) subsets of the form $\ps{\delta(\P)}_V=\alpha(\dastoo{V}{P})\subseteq\Sig_V$, one subset for each context $V\in\VN$. These subsets fit together to form a subobject $\ps{\delta(\P)}$ of $\Sig$. In our topos form of quantum logic, this subobject is the representative of the proposition ``$A\epsilon\Delta$''.

The condition for a collection $(\ps S_V)_{V\in\VN}$ of subsets of the Gel'fand spectra $\Sig_V,\;V\in\VN,$ to form a subobject $\ps S$ of $\Sig$ is the following: whenever one context $V'$ is contained in a larger context $V$, then the set $\ps S_V|_{V'}=\{\l|_{V'}\mid\l\in\ps S_V\}$ is contained in $\ps S_{V'}$. A subobject of $\Sig$ is nothing but a sub-presheaf of $\Sig$.

For clopen subsets, we can use equation (\ref{IsomP(V)Subcl(Sigma_V)}) to change from the clopen subsets $\ps S_V,\;V\in\VN,$ to the corresponding projections $\P_{\ps S_V},\;V\in\VN$. Clearly, $\P_{\ps S_V}\in\PV$ for all $V\in\VN$. On the level of projections, the condition for being a subobject simply becomes
\begin{equation}			\label{EqP_S_VLeqP_S_VPrime}
			\P_{\ps S_V}\leq\P_{\ps S_{V'}}
\end{equation}
for all $V',V\in\VN$ such that $V'\subset V$.

The subobjects of $\Sig$ that we construct from daseinisation all have the additional property that for each $V\in\VN$, we have a \emph{clopen} subset of $\Sig_V$. The idea is to consider subobjects with this property as the analogues of \emph{measurable} subsets of state space. We call these subobjects \emph{clopen subobjects}. We remark that there are many clopen subobjects $\ps S$ of $\Sig$ that do not come from daseinisation of a projection. It can be shown that the clopen subobjects form a complete Heyting algebra $\Subcl{\Sig}$, see thm. 14. in ref. \cite{DI08}. The meet (minimum, `And') $\ps S_1\wedge\ps S_2$ of two clopen subobjects $\ps S_1,\ps S_2\in\Subcl{\Sig}$ is defined stagewise:
\begin{equation}
			\ps S_1\wedge\ps S_2:=(\ps S_{1;V}\cap\ps S_{2;V})_{V\in\VN}.
\end{equation}
Here, $\ps S_{1;V}\cap\ps S_{2;V}$ simply is the intersection of the two clopen subsets $\ps S_{1;V},\ps S_{2;V}\in\Subcl{\Sig_V}$. Care must be taken when defining the meet of an infinite family of clopen subobjects, since for each $V$, the intersection of infinitely many clopen subsets need not be open. Hence, one has to take the interior of the set-theoretic intersection in order to obtain a clopen subset.

Similarly, the join (maximum, `Or') $\ps S_1\vee\ps S_2$ of two clopen subobjects is defined stagewise:
\begin{equation}
			\ps S_1\vee\ps S_2:=(\ps S_{1;V}\cup\ps S_{2;V})_{V\in\VN},
\end{equation}
where $\ps S_{1;V}\cup\ps S_{2;V}$ is the union of the two clopen sets $\ps S_{1;V},\ps S_{2;V}\in\Subcl{\Sig_V}$. For infinite joins, one has to take the closure of the set-theoretic union of clopen subsets at each stage $V$.

In contrast to meet and join, the negation $\neg\ps S$ of a clopen subobject $\ps S$ is not defined stagewise. Details can be found in ref. \cite{DI08}.

Heyting algebras are the algebraic structures which represent propositional intuitionistic logic. We have mapped our propositions (like ``$A\epsilon\Delta$'') about the physical world to the Heyting algebra $\Subcl{\Sig}$, which is given as a structure within our topos $\SetVNop$ of presheaves that is associated to the quantum system. Many further developments are possible, \cite{DI08} but here we just list a few properties of the daseinisation mapping:
\begin{itemize}
	\item If $\P<\hat Q$, then $\ps{\delta(\P)}<\ps{\delta(\hat Q)}$;
	\item the mapping $\ps{\delta}:\PN\rightarrow\Subcl{\Sig}$ is injective, but not surjective;
	\item $\ps{\delta(\hat 0)}=\ps 0$, the empty subobject, and $\ps{\delta(\hat 1)}=\Sig$;
	\item for all $\P,\hat Q\in\PN$, it holds that $\ps{\delta(\P\vee\hat Q)}=\ps{\delta(\P)}\vee\ps{\delta(\hat Q)}$;
	\item for all $\P,\hat Q\in\PN$, it holds that $\ps{\delta(\P\wedge\hat Q)}\leq\ps{\delta(\P)}\wedge\ps{\delta(\hat Q)}$. In general, $\ps{\delta(\P)}\wedge\ps{\delta(\hat Q)}$ is not of the form $\ps{\delta(\hat R)}$ for a projection $\hat R\in\PN$.
\end{itemize}
Daseinisation is a `translation' mapping between ordinary, Birkhoff-von Neumann quantum logic \cite{BvN36}, which is based upon the non-distributive lattice of projections $\PN$ in $\N$, and the topos form of propositional quantum logic, which is based upon the distributive lattice $\Subcl{\Sig}$, which more precisely is a Heyting algebra.


\section{Measures on the spectral presheaf}			\label{SecMeasures}
\subsection{Introduction}
We now want to consider measures on the spectral presheaf $\Sig$. The intuitive idea is that such a measure gives the `size' or `weight' of each subobject of $\Sig$ with respect to the measure, or at least the size of each subobject in a certain collection of `measurable' subobjects. In fact, we saw that the clopen subobjects of $\Sig$ play a special role, and we will measure their sizes. We emphasise that picking the complete Heyting algebra $\Subcl{\Sig}$ of clopen subobjects is a practical decision which allows us to prove the desired results, but leaves open a number of interesting questions on the relation between measurable sets and the spectral presheaf $\Sig$ as a \emph{topological} space in the topos $\SetVNop$. In future developments, one probably will consider a larger collection of subobjects than just the clopen ones. An additional remark: $\Subcl{\Sig}$ is a complete lattice, but not a Boolean algebra, hence it is not a $\sigma$-algebra, either. We very briefly discuss how measures behave with respect to complements, i.e., negation in the Heyting algebra $\Subcl{\Sig}$.

The measures we will define are probability measures, i.e., the whole of $\Sig$ will be of measure $1$ (in a suitable sense). Of course, the empty subobject will be of measure $0$. This section is partly motivated by the work by Jackson \cite{Jac06} on measure theory in topoi of sheaves. It also relates to the constructive measure theory by Coquand and Spitters \cite{CS08} and its application to topoi associated to quantum systems.\cite{HLS08} We will present results with respect to the topos of presheaves $\SetVNop$ that is used in the topos formulation of quantum theory \cite{DI(1),DI(2),DI(3),DI(4),Doe07b,DI08}. Our arguments are topos-external, which allows for very concrete reasoning without the intricacies of constructive mathematics. The main results are an abstract characterisation of measures on clopen subobjects and the proof that there exists a bijection between measures and states of the quantum system. In particular, given a measure $\mu$ on the clopen subobjects of the spectral presheaf, one can reconstruct a state $\rho_\mu$ of the quantum system from it.


\subsection{Definition of a measure from a quantum state}
In classical physics, a general state of a physical system is given mathematically as a probability measure on the state space of the system, while Dirac measures represent `pure' states. Usually, the Dirac measure $\delta(s)$ concentrated at a point $s$ of state space is identified with the point itself. A probability measure assigns a number between $0$ and $1$ to each measurable subset of state space, and this number can be interpreted as the size or weight of the subset with respect to the measure.

For a quantum system, as we saw, the spectral presheaf $\Sig$ belonging to the system takes the role of a state object (but, being a presheaf, $\Sig$ is not a space or a set). In analogy to the classical case, we expect that every state of the quantum system gives us a measure on (clopen subobjects of) the spectral presheaf $\Sig$. Here, a state of the quantum system is given by a state $\rho$ of the von Neumann algebra $\N$ of physical quantities, i.e., a positive linear functional $\rho:\N\rightarrow\mathbb{C}$ of norm $1$. This notion of (mathematical) states comprises the usual notion of physical states, including vector states and density matrices. In physics, attention is often restricted to \emph{normal} states, \cite{KR83} which are those corresponding to (finite or infinite) density matrices. We will characterise those measures that correspond to normal states. On the other hand, there are von Neumann algebras that do not possess any normal states (namely algebras of type $III$), but which do have physically important states like KMS states. The results we present apply to these algebras and states as well. As a consequence, the measures we define are finitely additive only in general, not $\sigma$-additive. (This deviates from the standard conventions in measure theory. Some readers may prefer to call our measures \emph{proto-measures} or similar.)

We will see shortly that the definition of a measure from a state works in a straightforward manner. A slightly unusual point is that the measure of a clopen subobject $\ps S$ is not a single real number between $0$ and $1$, but rather a collection of such numbers, one for each context $V\in\VN$. This of course comes from the fact that a clopen subobject $\ps S$ of $\Sig$ is not simply a set, but consists of a collection of sets, namely one set $\ps S_V$ for each context $V\in\VN$. Moreover, whenever we have $V'\subset V$, then there is a function from $\ps S_V$ to $\ps S_{V'}$, given by restriction of the presheaf $\Sig$ (see section \ref{SecProposAsSubobjs}). The measures we define are monotone in the following sense: the smaller the context, the larger the number assigned to it. The definition of a measure is as follows:

Let $\rho:\N\rightarrow\mathbb{C}$ be a state of the algebra $\N$ of physical quantities of the quantum system, i.e., a positive linear functional of norm $1$. Then the \emph{measure $\mu_\rho$ on clopen subobjects associated to $\rho$} is the mapping
\begin{eqnarray}			\label{Defmu_rho}
			\mu_\rho:\Subcl{\Sig} &\longrightarrow& \Gamma\ps{[0,1]^\succeq}\\	\nonumber	
			\ps S=(\ps S_V)_{V\in\VN} &\longmapsto& (\rho(\P_{\ps S_V}))_{V\in\VN}.
\end{eqnarray}
Given a clopen subobject $\ps S$, we assign the real number $\rho(\P_{\ps S_V})\in [0,1]$, that is, the expectation value of the projection $\P_{\ps S_V}$ in the state $\rho$, to each context $V$. Here, $\P_{\ps V}=\alpha^{-1}(\ps S_V)$ is the projection in $V$ corresponding to the clopen subset $\ps S_V\subseteq\Sig_V$ via the isomorphism (\ref{IsomP(V)Subcl(Sigma_V)}). If we have two contexts $V',V$ such that $V'\subset V$, then $\P_{\ps S_{V'}}\geq\P_{\ps S_V}$ (see (\ref{EqP_S_VLeqP_S_VPrime})), so $\rho(\P_{\ps S_{V'}})\geq\rho(\P_{\ps S_V})$. For each clopen subobject $\ps S$ of $\Sig$, we thus obtain a function $\mu_\rho(\ps S):\VN\rightarrow [0,1]$, given by $\mu_\rho(\ps S)(V)=\rho(\P_{\ps S_V})$. This function is order-reversing as a function from the partially ordered set $\VN$ to the interval $[0,1]$, equipped with the usual order coming from the real numbers. Such functions can be regarded as global elements of a presheaf $\ps{[0,1]^\succeq}$, which explains the notation for the codomain of $\mu_\rho$ in (\ref{Defmu_rho}). The presheaf $\ps{[0,1]^\succeq}$ is a sub-presheaf of the presheaf $\ps{\mathbb{R}^\succeq}$ that is extensively discussed in refs. \cite{DI(3),DI08}.

Order-reversing functions from $\VN$ to $[0,1]$ can be added, and addition is defined stage by stage. In particular, let $\ps S\in\Subcl{\Sig}$, and let $\mu_{\rho_1}(\ps S),\mu_{\rho_2}(\ps S):\VN\rightarrow [0,1]$ be two such functions, then, for all $V\in\VN$,
\begin{equation}
			(\mu_{\rho_1}+\mu_{\rho_2})(\ps S)(V):=\mu_{\rho_1}(\ps S)(V)+\mu_{\rho_2}(\ps S)(V).
\end{equation}
The sum $\mu_{\rho_1}(\ps S)+\mu_{\rho_2}(\ps S)$ is a real-valued, order-reversing function on $\VN$, but it need not necessarily take values in the interval $[0,1]$. By taking a convex combination of measures $\mu_{\rho_1},\mu_{\rho_2}$, given for all $\ps S\in\Subcl{\Sig}$ and all $V\in\VN$ by
\begin{equation}
			(c\mu_{\rho_1}+(1-c)\mu_{\rho_2})(\ps S)(V):=c\mu_{\rho_1}(\ps S)(V)+(1-c)\mu_{\rho_2}(\ps S)(V),
\end{equation}
one obtains a new measure $c\mu_{\rho_1}+(1-c)\mu_{\rho_2}$ on clopen subobjects with values in $\Gamma\ps{[0,1]^\succeq}$, and, clearly, this is the measure $\mu_{c\rho_1+(1-c)\rho_2}$ associated to the state $c\rho_1+(1-c)\rho_2$ of $\N$. The space of measures of the form
\begin{equation}
			\mu_\rho:\Subcl{\Sig}\longrightarrow\Gamma\ps{[0,1]^\succeq},
\end{equation}
hence is a convex space, and the convex structure is exactly mirroring the convex structure on the space $\mathcal{S}(\N)$ of states of the von Neumann algebra $\N$.


\subsection{Properties of the measure $\mu_\rho$}
Let a (fixed) state $\rho$ and its associated measure $\mu_\rho$ be given. We will show some properties of $\mu_\rho$ that justify it being called a measure.

As before, let $\ps 0$ denote the empty subobject of $\Sig$. For each context $V\in\VN$, we have $\ps 0_V=\emptyset$, and the empty set in the Gel'fand spectrum $\Sig_V$ corresponds to the null projection $\hat 0$ in the projection lattice $\mathcal{P}(V)$ via the lattice isomorphism (\ref{IsomP(V)Subcl(Sigma_V)}). We obtain, for all $V$,
\begin{equation}
			\mu_\rho(\ps 0)(V)=\rho(\hat 0)=0,
\end{equation}
so globally
\begin{equation}
			\mu_\rho(\ps 0)=0_\VN
\end{equation}
where the right hand side denotes the collection of numbers $0$, one for each context $V\in\VN$. This is the global element of $\ps{[0,1]^\succeq}$ that is constantly $0$. In a similar manner, one gets
\begin{equation}
			\mu_\rho(\Sig)=1_\VN,
\end{equation}
where the right hand side denotes the collection of numbers $1$, one for each context $V$. This is the global element of $\ps{[0,1]^\succeq}$ that is constantly $1$. As expected, the whole of $\Sig$ is of measure $1$ in the appropriate sense, and the empty subobject is of measure $0$.

Let $\ps S_1,\ps S_2$ be two disjoint clopen subobjects of $\Sig$, i.e., within the Heyting algebra $\Subcl{\Sig}$, we have $\ps S_1\wedge\ps S_2=\ps 0$. The union $\ps S_1\vee\ps S_2$ of two clopen subobjects is given by the stagewise operations $(\ps S_1\vee\ps S_2)_V=\ps S_{1;V}\cup\ps S_{2;V}$ for all $V$. We obtain
\begin{eqnarray*}
			\mu_\rho(\ps S_1\vee\ps S_2)(V) &=& \mu_\rho((\ps S_1\vee\ps S_2)_V)\\
			&=& \mu_\rho(\ps S_{1;V}\cup\ps S_{2;V})\\
			&=& \rho(\P_{\ps S_{1;V}}\vee\P_{\ps S_{2;V}})\\
			&=& \rho(\P_{\ps S_{1;V}}+\P_{\ps S_{2;V}})\\
			&=& \mu_\rho(\ps S_1)(V)+\mu_\rho(\ps S_2)(V)
\end{eqnarray*}
for all $V\in\VN$ and hence globally
\begin{equation}			\label{FinAdditivity}
			\mu_\rho(\ps S_1\vee\ps S_2)=\mu_\rho(\ps S_1)+\mu_\rho(\ps S_2).
\end{equation}
This is another basic property one expects from a measure, namely that it behaves additively on disjoint subsets (here, subobjects). Of course, equation (\ref{FinAdditivity}) generalises to arbitrary finite collections of pairwise disjoint subobjects $\ps S_1,...,\ps S_n$. Every measure $\mu_\rho$ hence is \emph{finitely additive}, which follows from the fact that the quantum state $\rho$ is linear.

In standard measure theory, it is often assumed that measures are not only finitely additive, but $\sigma$-additive: if $\mu$ is a measure on a collection $\mathcal{M}(X)$ of (measurable) subsets of some space $X$, then for arbitrary countable families $(S_i)_{i\in I}$ of pairwise disjoint sets in $\mathcal{M}(X)$, one has
\begin{equation}
			\mu(\bigcup_{i\in I}S_i)=\sum_{i\in I}\mu(S_i).
\end{equation}
Some measures behave additively on collections $(S_j)_{j\in J}$ of arbitrary cardinality. Such measures are simply called \emph{additive}.

Since the clopen subobjects of $\Sig$ are not sets, but presheaves, some care must be taken when considering generalisations of these properties. Even for a two-dimensional Hilbert space $\mathbb{C}^2$ and the von Neumann algebra $\mathcal{B}(\mathbb{C}^2)$, there exist infinitely many contexts $V\in\mathcal{V}(\mathcal{B}(\mathbb{C}^2))$. (Some simple arguments show that the space of contexts $\mathcal{V}(\mathcal{B}(\mathbb{C}^2))$ is isomorphic to the projective space $P\mathbb{R}^2$.) For each $V$, define a clopen subobject $\ps S^V$ of $\Sig$ by setting $(\ps S^V)_V=\{\l_1\}$, where $\l_1$ is one element (picked arbitrarily from the two elements) of the Gel'fand spectrum $\Sig_V$ of $V$, and $(\ps S^V)_{\tilde V}=\emptyset$ for all $\tilde V\neq V$. The family $(\ps S^V)_{V\in\mathcal{V}(\mathcal{B}(\mathbb{C}^2))}$ of pairwise disjoint, clopen subobjects of $\Sig$ is uncountable, but trivially, for any quantum state $\rho$, one obtains
\begin{equation}
			\mu_\rho(\bigvee_{V\in\mathcal{V}(\mathcal{B}(\mathbb{C}^2))}\ps S^V)=
			\sum_{V\in\mathcal{V}(\mathcal{B}(\mathbb{C}^2))}\mu_\rho(\ps S^V).
\end{equation}
Similar arguments can be made for other algebras of the form $\mathcal{B}(\mathbb{C}^n)$ and more generally for any non-commutative von Neumann algebra $\N$. In this trivial sense, every measure $\mu_\rho$ is additive for certain uncountable families of pairwise disjoint families of clopen subobjects.

On the other hand, one may want to consider a countably infinite family $(\ps S_i)_{i\in I}$ of clopen subobjects such that, for a suitable context $V$, one has that the clopen subsets $\ps S_{i;V},\;i\in I,$ of $\Sig_V$ are pairwise disjoint. Let $\P_i,\;i\in I,$ denote the countably infinite family of pairwise orthogonal projections in $\mathcal{P}(V)$ corresponding to the sets $\ps S_{i;V},\;i\in I$. Then, locally at $V$, we consider
\begin{equation}
			(\mu_\rho(\bigvee_{i\in I}\ps S_i))(V)=\rho(\bigvee_{i\in I}\P_i),
\end{equation}
which, in general, is only equal to 
\begin{equation}
			\sum_{i\in I}\mu_\rho(\ps S_i)(V)=\sum_{i\in I}\rho(\P_i)
\end{equation}
if $\rho$ is a \emph{normal} state of $\N$. This is the connection between $\sigma$-additivity and normal states that one might expect. The caveat is that the subobjects $(\ps S_i)_{i\in I}$ are only pairwise disjoint locally at $V$ (i.e., the clopen subsets $\ps S_{i;V},\;i\in I,$ of $\Sig_V$ are pairwise disjoint), but globally, the subobjects $(\ps S_i)_{i\in I}$ are not pairwise disjoint. This can easily be seen: for definiteness, assume that the countable index set $I$ contains the number $1$ (e.g., $I=\mathbb{N}$). Consider the context $V_1:=\{\P_1,\hat 1\}''$. Then, from the properties of clopen subobjects of $\Sig$ (specifically, as a consequence of formula (\ref{EqP_S_VLeqP_S_VPrime})), one has $\ps S_{i;V_1}\supseteq\{\l_{\hat 1-\P_1}\}$ for all $i\in I,\;i\neq 1$, where $\l_{\hat 1-\P_1}$ is the element of the Gel'fand spectrum $\Sig_{V_1}$ of $V_1$ that sends $\hat 1-\P_1$ to $1$ and $\P_1$ to $0$. This means that at $V_1$, all the subobjects $(\ps S_i)_{i\in I\backslash\{1\}}$ overlap and hence are not disjoint.

Formula (\ref{FinAdditivity}), finite additivity, is a special case of the following result: let $\ps S_1,\ps S_2$ be two arbitrary clopen subobjects of $\Sig$. Then, for all $V\in\VN$,
\begin{eqnarray*}
			& &\mu_\rho(\ps S_1\vee\ps S_2)(V)+\mu_\rho(\ps S_1\wedge\ps S_2)(V)\\
			&=& \rho(\P_{\ps S_{1;V}}\vee\P_{\ps S_{2;V}})+\rho(\P_{\ps S_{1;V}}\wedge\P_{\ps S_{2;V}})\\
			&=& \rho(\P_{\ps S_{1;V}}\vee\P_{\ps S_{2;V}}+\P_{\ps S_{1;V}}\wedge\P_{\ps S_{2;V}})\\
			&=& \rho(\P_{\ps S_{1;V}}+\P_{\ps S_{2;V}})\\
			&=& \rho(\P_{\ps S_{1;V}})+\rho(\P_{\ps S_{2;V}})\\
			&=& \mu_\rho(\ps S_1)(V)+\mu_\rho(\ps S_2)(V),
\end{eqnarray*}
which gives globally
\begin{equation}
			\mu_\rho(\ps S_1\vee\ps S_2)+\mu_\rho(\ps S_1\wedge\ps S_2)=\mu_\rho(\ps S_1)+\mu_\rho(\ps S_2).
\end{equation}
A property of this form is characteristic for measures in general, and $\mu_\rho$ fulfils it for clopen subobjects of our quantum state space analogue, the spectral presheaf $\Sig$.

When considering negation of clopen subobjects, the fact that $\Subcl{\Sig}$ is a Heyting algebra and not a Boolean algebra plays a role. Let $\ps S$ be a clopen subobject, and let $\neg\ps S$ be its negation in the Heyting algebra $\Subcl{\Sig}$. It is easy to see that if $\ps S$ is neither the whole of $\Sig$ nor the empty subobject $\ps 0$, then the strict inequality
\begin{equation}
			\ps S\vee\neg\ps S<\Sig
\end{equation}
holds. $\neg\ps S$ is the largest clopen subobject of $\Sig$ that is disjoint from $\ps S$, but since we have a Heyting algebra, $\neg\ps S$ is only a pseudo-complement, not an actual complement in general. For our measure $\mu_\rho$, this implies
\begin{equation}
			\mu_\rho(\ps S\vee\neg\ps S)<1_\VN
\end{equation}
for non-trivial clopen subobjects $\ps S$. This is in contrast to ordinary measure theory, where the union of a measurable set $S$ and its complement $S^c$ is the whole measure space $X$ (and hence, if one considers probability measures, is of measure $1$).


\subsection{Abstract characterisation of measures and reconstruction of states}
We now axiomatise the notion of a measure: let a quantum system with corresponding non-commutative von Neumann algebra  $\N$ of physical quantities and spectral presheaf $\Sig$ be given. A mapping
\begin{eqnarray}			\label{Defmu}
			\mu:\Subcl{\Sig} &\longrightarrow& \Gamma\ps{[0,1]^\succeq}\\
			\ps S=(\ps S_V)_{V\in\VN} &\longmapsto& (\mu(\ps S_V))_{V\in\VN}
\end{eqnarray}
is called a \emph{measure on the clopen subobjects of $\Sig$} if the following two conditions are fulfilled:
\begin{enumerate}
	\item $\mu(\Sig)=1_\VN$;
	\item for all $\ps S_1,\ps S_2\in\Subcl{\Sig}$, it holds that\newline $\mu(\ps S_1\vee\ps S_2)+\mu(\ps S_1\wedge\ps S_2) =\mu(\ps S_1)+\mu(\ps S_2)$.
\end{enumerate}
In particular, this definition means that for each clopen subobject $\ps S$ of $\Sig$, we obtain a global element $\mu(\ps S)$ of $\ps{[0,1]^\succeq}$. Such a global element is an order-reversing function $\mu(\ps S):\VN\rightarrow [0,1]$ from the partially ordered set of contexts to the unit interval. By definition (\ref{Defmu}), we have $\mu(\ps S)(V)=\mu(\ps S_V)$ for all $V\in\VN$. In particular, $\mu(\ps S)$ is local in the sense that $\mu(\ps S)(V)$ only depends on $\ps S_V$, but not on any other components $\ps S_{\tilde V}$ for $\tilde V\neq V$. If $V,V'$ are two contexts such that $V'\subset V$, then $0\leq\mu(\ps S)(V)\leq\mu(\ps S)(V')\leq 1$. It is easy to see from properties (1.) and (2.) of a measure $\mu$ that
\begin{equation}			\label{mu(0)=0}
			\mu(\ps 0)=0_{\VN}.
\end{equation}

We saw that every state $\rho$ of the von Neumann algebra $\N$ defines a measure $\mu_\rho$ as in (\ref{Defmu_rho}). The main result of this paper is that, conversely, if $\N$ has no direct summand of type $I_2$, then every measure $\mu$ determines a unique state $\rho_\mu$ such that we obtain a bijection between $\mathcal{S}(\N)$, the state space of the von Neumann algebra $\N$, and $\mathcal{M}(\Subcl{\Sig})$, the set of measures on the clopen subobjects of the spectral presheaf $\Sig$.

The idea of the proof is to show that a measure $\mu$ determines a unique mapping $m:\PN\rightarrow [0,1]$ from the projection operators in the von Neumann algebra $\N$ to the unit interval such that (i) for the identity operator $\hat 1$, we have $m(\hat 1)=1$ and (ii) if $\P,\hat Q$ are orthogonal projections, then $m(\P\vee\hat Q)=m(\P+\hat Q)=m(\P)+m(\hat Q)$. Such a mapping $m$ is called a \emph{finitely additive probability measure on the projections of $\N$}. From the generalised version of Gleason's theorem \cite{Mae89} (the reference gives a detailed review of this strong result and its proof), it is known that each such finitely additive probability measure on the projections determines a unique state of $\N$, provided that $\N$ contains no direct summand of type $I_2$. Hence, from a measure $\mu$ on clopen subobjects of $\Sig$, we will obtain a unique finitely additive measure $m$ on the projections of $\Sig$ and from this a unique state $\rho_\mu$ of $\N$.

Let $\mu$ be a measure on the clopen subobjects of $\Sig$. We define $m(\hat 1):=1$, which must be fulfilled by any finitely additive probability measure on the projections and, moreover, is justified by property (1.) of the measure $\mu$.

Now let $\P\in\PN$ be any projection, and let $\ps S$ be a clopen subobject of $\Sig$ such that for some context $V$, we have that the clopen subset $\ps S_V\subseteq\Sig_V$ corresponds to the projection $\P$ via (\ref{IsomP(V)Subcl(Sigma_V)}). We will constantly use this lattice isomorphism between clopen subsets in the Gel'fand spectrum of a context $V$ and the projections in $V$. We define
\begin{equation}			\label{Defm(P)}
			m(\P):=\mu(\ps S)(V)=\mu(\ps S_V).
\end{equation}
Of course, we have to show that this does not depend on the choice of the subobject $\ps S$ and the context $V$. The potential difficulty is the following: $\ps S_V\subseteq\Sig_V$ is a clopen subset of some Gel'fand spectrum which corresponds to the projection $\P\in\PN$. There are many other clopen subsets of other Gel'fand spectra that also correspond to the same projection $\P$. These subsets may show up as components $\ps{\tilde S}_{\tilde V}$ of other subobjects $\ps{\tilde S}$ at other contexts $\tilde V$. Since these clopen subsets lie in different Gel'fand spectra, they cannot be compared directly (while the corresponding projection $\P$ is the same). The task is to show that the measure $\mu$, which acts on the level of clopen subobjects of $\Sig$ and locally on the level of clopen subsets of Gel'fand spectra, will assign the same number to all these subsets. I.e., we have to show that $\mu(\ps S)(V)=\mu(\ps{\tilde S})(\tilde V)$ whenever $\ps S_V$ and $\ps{\tilde S}_{\tilde V}$ both correspond to the same projection $\P$. 

First, let us keep $\ps S$ fixed, and assume that there is a smaller context $V'\subset V$ such that $\P$ is contained in both $V$ and $V'$. Since $\hat 1$ is contained in every context, we also have $\hat 1-\P\in V',V$. Moreover, assume that $\ps S_{V'}$ is such that the projection corresponding to this clopen subset of $\Sig_{V'}$ is $\P$. (For example, the subobject $\ps{\delta(\P)}$ has this property.) We have to show that $\mu(\ps S)(V)=\mu(\ps S)(V')$. Let $\ps{S^c}$ be another clopen subobject such that both $\ps{S^c}_V\subseteq\Sig_V$ and $\ps{S^c}_{V'}\subseteq\Sig_{V'}$ correspond to $\hat 1-\P$. ($\ps{\delta(\hat 1-\P)}$ has this property.) Then we have
\begin{eqnarray}			\label{DisjointAtVAndVPrime}
			(\ps S\wedge\ps{S^c})_V=\ps 0_V=\emptyset &,& \ \ (\ps S\wedge\ps{S^c})_{V'}=\ps 0_{V'}=\emptyset,\\
						\label{JoinGivesSigAtVAndVPrime}
			(\ps S\vee\ps{S^c})_V=\Sig_V &,& \ \ (\ps S\vee\ps{S^c})_{V'}=\Sig_{V'}.
\end{eqnarray}
From properties (1.) and (2.) of the measure $\mu$, we obtain
\begin{eqnarray*}
			& &\mu(\Sig)(V)=1\\
			&=& \mu(\ps S\vee\ps{S^c})(V)\\
			&=& \mu(\ps S)(V)+\mu(\ps{S^c})(V)-\mu(\ps S\wedge\ps{S^c})(V).
\end{eqnarray*}
The last term is $0$ from equation (\ref{mu(0)=0}), so we obtain
\begin{equation}
			\mu(\ps S)(V)+\mu(\ps{S^c})(V)=1,
\end{equation}
and similarly
\begin{equation}
			\mu(\ps S)(V')+\mu(\ps{S^c})(V')=1.
\end{equation}
Since $\mu(\ps S):\VN\rightarrow [0,1]$ is an order-reversing function, we also have
\begin{eqnarray}
			& &\mu(\ps S)(V')\geq\mu(\ps S)(V),\\
			& &\mu(\ps{S^c})(V')\geq\mu(\ps{S^c})(V).
\end{eqnarray}
Taken together, this implies
\begin{eqnarray}			\label{ConstForDiffContexts}
			& &\mu(\ps S)(V')=\mu(\ps S)(V),\\
			& &\mu(\ps{S^c})(V')=\mu(\ps{S^c})(V)=1-\mu(\ps S)(V).
\end{eqnarray}
We have now shown (equation (\ref{ConstForDiffContexts})) that if $\ps S$ is a fixed clopen subobject and $V',V$ are two contexts such that $V'\subset V$ and both $\ps S_V$ and $\ps S_{V'}$ correspond to the same projection, then $\mu$ assigns the same number to them.

In our proof so far, we assumed that both $\ps S_V$ and $\ps S_{V'}$ correspond to the same projection $\P$. While this is the situation we are interested in, it also is an additional piece of information that cannot be read off from the subobject $\ps S$ directly. We want to point out that the existence of another subobject $\ps{S^c}$ such that (\ref{DisjointAtVAndVPrime}) and (\ref{JoinGivesSigAtVAndVPrime}) hold suffices for the proof to go through. Hence, the argument can be made on the level of subobjects, without direct reference to the projection $\P$.

We need another small observation: if $\ps S$ and $\ps{\tilde S}$ are two subobjects that coincide at $V$, i.e., $\ps S_V=\ps{\tilde S}_V$, then we obtain from property (2.) of a measure $\mu$
\begin{eqnarray*}
			& &\mu(\ps S)(V)+\mu(\ps{\tilde S})(V)\\
			&=& \mu(\ps S\vee\ps{\tilde S})(V)+\mu(\ps S\wedge\ps{\tilde S})(V)\\
			&=& \mu((\ps S\vee\ps{\tilde S})_V)+\mu((\ps S\wedge\ps{\tilde S})_V)\\
			&=& \mu(\ps S_V\cup\ps{\tilde S}_V)+\mu(\ps S_V\cap\ps{\tilde S}_V)\\
			&=& \mu(\ps S_V)+\mu(\ps S_V),\\
			&=& \mu(\ps S)(V)+\mu(\ps S)(V),
\end{eqnarray*} 
which implies
\begin{equation}			\label{SameProjSameValue}
			\mu(\ps S)(V)=\mu(\ps{\tilde S})(V).
\end{equation}
Again, this argument does not refer to the projection corresponding to the components $\ps S_V,\ps{\tilde S}_V$. 

To complete the proof that $m(\P)$ is well defined by (\ref{Defm(P)}), we of course need to refer to $\P$ explicitly: we assume as before that $\ps S$ is a clopen subobject such that $\ps S_V$ corresponds to $\P$, and let $\ps{\tilde S}$ be another clopen subobject and $\tilde V$ another context such that $\ps{\tilde S}_{\tilde V}$ also corresponds to $\P$. In particular, both contexts $V,\tilde V$ contain the projection $\P$. Then the context $V\cap\tilde V$ also contains $\P$. The subobject $\ps{\delta(\P)}$ coincides with $\ps S$ at $V$, i.e., $\ps{\delta(\P)}_V=\ps S_V$, and it coincides with $\ps{\tilde S}$ at $\tilde V$, i.e., $\ps{\delta(\P)}_{\tilde V}=\ps{\tilde S}_{\tilde V}$. Moreover, the clopen subset $\ps{\delta(\P)}_{V\cap\tilde V}\subseteq\Sig_{V\cap\tilde V}$ corresponds to the projection $\P$ (while $\ps S_{V\cap\tilde V}$ and $\ps{\tilde S}_{V\cap\tilde V}$ need not necessarily correspond to $\P$). We obtain
\begin{eqnarray*}
			\mu(\ps S)(V)&\overset{(\ref{SameProjSameValue})}=& \mu(\ps{\delta(\P)})(V)\\
			&\overset{(\ref{ConstForDiffContexts})}=& \mu(\ps{\delta(\P)})(V\cap\tilde V)\\
			&\overset{(\ref{ConstForDiffContexts})}=& \mu(\ps{\delta(\P)})(\tilde V)\\
			&\overset{(\ref{SameProjSameValue})}=& \mu(\ps{\tilde S})(\tilde V).
\end{eqnarray*}
This shows that the value $m(\P)=\mu(\ps S)(V)$ in (\ref{Defm(P)}) is well-defined.

Let $\P,\hat Q$ be two orthogonal projections, let $V\in\VN$ be a context that contains both $\P$ and $\hat Q$, let $\ps{S^\P}$ be a subobject such that the clopen subset $\ps{S^\P}_V\subseteq\Sig_V$ corresponds to $\P$, and let $\ps{ S^{\hat Q}}$ be another subobject such that $\ps{S^{\hat Q}}_V$ corresponds to $\hat Q$. Then $(\ps{S^\P}\vee\ps{S^{\hat Q}})_V$ corresponds to $\P\vee\hat Q$ and we obtain
\begin{eqnarray*}
			& &m(\P\vee\hat Q)=\mu(\ps{S^\P}\vee\ps{S^{\hat Q}})(V)\\
			&=& \mu(\ps{S^\P})(V)+\mu(\ps{S^{\hat Q}})(V)+\mu(\ps{S^\P}\wedge\ps{S^{\hat Q}})(V)\\
			&=& \mu(\ps{S^\P})(V)+\mu(\ps{S^{\hat Q}})(V)\\
			&=& m(\P)+m(\hat Q).
\end{eqnarray*}
This shows that $m:\PN\rightarrow [0,1]$ actually is finitely additive. $m(\hat 0)=0$ follows easily from this and $m(\hat 1)=1$.

Summing up, we have shown that each measure $\mu$ on the clopen subobjects of $\Sig$ defines a unique finitely additive measure $m$ on the projections of $\N$. From the generalised version of Gleason's theorem, \cite{Mae89} we know that $m$ extends to a unique state $\rho_\mu$ of the von Neumann algebra $\N$, provided $\N$ has no summand of type $I_2$. Obviously, the mappings $\rho\mapsto\mu_\rho$ and $\mu\mapsto\rho_\mu$ are inverse to each other.

\begin{theorem}			\label{ThmStatesMeasures}
For every von Neumann algebra $\N$ with no direct summand of type $I_2$, there exists a bijection between the space $\S(\N)$ of states of $\N$ and the set $\mathcal{M}(\Subcl{\Sig})$ of measures on the clopen subobjects of the spectral presheaf $\Sig$ belonging to $\N$. The mapping from a state $\rho$ to a measure $\mu_\rho$ is given by (\ref{Defmu_rho}), and the mapping from a measure $\mu$ to a state $\rho_\mu$ is given as the composite $\mu\mapsto m\mapsto\rho_\mu$ as described above. In particular, the set $\mathcal{M}(\Subcl{\Sig})$ of measures is a convex space.
\end{theorem}

A measure $\mu$ on the clopen subobjects is called \emph{locally $\sigma$-additive} if it has the property that for all countable families $(\ps S_i)_{i\in I}$ of clopen subobjects that are locally disjoint at some context $V\in\VN$ (i.e., the clopen subsets $(\ps S_{i;V}),\;i\in I,$ of $\Sig_V$ are pairwise disjoint), it holds that
\begin{equation}
			\mu(\bigvee_{i\in I}\ps S_i)(V)=\mu(\bigvee_{i\in I}\ps S_{i;V})=\sum_{i\in I}\mu(\ps S_{i;V})=\sum_{i\in I}\mu(\ps S_i)(V).
\end{equation}
Let $m$ be the measure on projections determined by a locally $\sigma$-additive measure $\mu$ on subobjects. It is clear by construction that $m$ is $\sigma$-additive on projections. The state $\rho_\mu$ obtained from $m$ by Gleason's theorem then is a \emph{normal} state. Conversely, every normal state determines a locally $\sigma$-additive measure.

\begin{corollary}
The normal states of a von Neumann algebra $\N$ with no type $I_2$-summand correspond to the locally $\sigma$-additive measures on clopen subobjects under the bijection described in Thm. \ref{ThmStatesMeasures}.
\end{corollary}


\section{Measures and expectation values}			\label{SecMeasExpVals}
In this section, we want to sketch the connections between measures on clopen subobjects of $\Sig$, pseudo-states as representatives of vector states as discussed in ref. \cite{DI08}, and expectation values in ordinary quantum theory.

\subsection{Pseudo-states and their reconstruction from measures}
Let $\psi$ be a unit vector in Hilbert space. It commonly is identified with the vector state
\begin{eqnarray}
			\langle\psi,\_\psi\rangle:\N &\longrightarrow& \mathbb{C}\\			\nonumber
			\A &\longmapsto& \langle\psi,\A\psi\rangle.
\end{eqnarray}
Let $\P^\psi$ denote the projection onto the one-dimensional subspace $\mathbb{C}\psi$ of Hilbert space determined by $\psi$. (This projection is often denoted as $|\psi\rangle\langle\psi|$, but this notation is a little clumsy for our purposes.) In the topos approach \cite{DI08}, a vector state like $\psi$ is represented by a so-called \emph{pseudo-state} $\wpsi$, given by
\begin{equation}			\label{DefPseudoState}
			\wpsi:=(\wpsi_V)_{V\in\VN}=\ps{\delta(\P^\psi)}.
\end{equation}
The pseudo-state $\wpsi$ hence is a clopen subobject of $\Sig$. The interpretation is that this subobject is the smallest subobject of $\Sig$ which represents a proposition that is totally true in the state $\psi$. This can be compared with the classical case: the analogue of a vector state $\psi$ is a `pure' state, represented by a Dirac measure $\delta(s)$. The propositions which are true in this state are represented by those measurable subsets of state space that contain the element $s$, i.e., those subsets that are of measure $1$ with respect to $\delta(s)$. The smallest such set is the one-element set $\{s\}$, i.e., a single point of state space (and this set is measurable with respect to the Dirac measure $\delta(s)$). Using this analogy, the pseudo-state $\wpsi\subset\Sig$ is what corresponds to the one-point subset $\{s\}$ of the classical state space.

It is an interesting fact that $\wpsi$ is \emph{not} a global element of the presheaf $\Sig$. Global elements of a presheaf are the category-theoretical analogues of points, but the spectral presheaf $\Sig$ does not have any points in this sense. This result is just the Kochen-Specker theorem \cite{IB98,IB00,Doe05,DI08}. In the sense described above, the pseudo-states $\wpsi$ are `as close to being points of $\Sig$ as possible'.

Definition (\ref{DefPseudoState}) can be rewritten slightly: for each $V\in\VN$, we have
\begin{equation}
			\wpsi_V=\ps{\delta(\P^\psi)}_V\overset{(\ref{DefDasOfProjs})}=\alpha(\delta^o(\P^\psi)_V).
\end{equation}
By definition (\ref{DefOuterDas}),
\begin{eqnarray}
			\delta^o(\P^\psi)_V &=& \bigwedge\{\hat Q\in\PV\mid\hat Q\geq\P^\psi\}\\			\nonumber
			&=& \bigwedge\{\hat Q\in\PV\mid\langle\psi,\hat Q\psi\rangle=1\}.
\end{eqnarray}
This is the smallest projection in $V$ that has expectation value $1$ with respect to the vector state $\psi$. Thus, for all $V\in\VN$,
\begin{equation}
			\wpsi_V=\alpha(\bigwedge\{\hat Q\in\PV\mid\langle\psi,\hat Q\psi\rangle=1\}).
\end{equation}
We now use definition (\ref{Defmu_rho}), applied to the state $\psi$, to obtain a measure $\mu_\psi$ on the clopen subobjects of $\Sig$. Clearly, the subobject $\wpsi=(\wpsi_V)_{V\in\VN}$ is of measure $1$ with respect to $\mu_\psi$ (i.e., $\mu_\psi(\wpsi)=1_{\VN}$), and there is no smaller clopen subobject that is of measure $1$ with repsect to $\mu_\psi$. This is in complete analogy with the classical case where the one-element set $\{s\}$ is the smallest subset of measure $1$ with respect to the Dirac measure $\delta(s)$.

We have shown that the pseudo-state $\wpsi$ can be reconstructed from the measure $\mu_\psi$, simply as the smallest subobject that is of measure $1$ with respect to $\mu_\psi$. It is a very interesting open question if, conversely, the measure $\mu_\psi$ can be constructed from the pseudo-state $\wpsi$. The latter suffices to assign topos-internal truth-values to all propositions of the form ``$A\epsilon\Delta$'', while the measure $\mu_\psi$ can deliver the usual expectation values of quantum theory for the vector state $\psi$. This is sketched below in subsection (\ref{SubsecMeasuresExpVals}) and will be developed in a future publication. 

In the classical case, the analogous construction of the measure from the minimal subset of measure $1$ is trivial for `pure' states: the Dirac measure $\delta(s)$ can be read off directly from the one-element set $\{s\}$. On the other hand, one cannot reconstruct a unique probability measure on the classical state space if the minimal subset of measure $1$ is larger than a one-element set. Analogously, we expect in the quantum case that we can at best construct $\mu_\psi$ uniquely from $\wpsi$, but not a more general measure $\mu$ from a minimal subobject of measure $1$ that is larger than some $\wpsi$.

The topos approach to quantum theory aims at a realist description in which measurements, statistics and probabilities play a secondary role only, while logical, non-instrumentalist aspects are central. For this reason, it would be highly satisfying if the measure $\mu_\psi$, which contains the probabilistic information, could be reconstructed from the pseudo-state $\wpsi$, which contains the logical aspects. If such a construction of the measure from the pseudo-state is possible, then the purely quantum-theoretical aspects of probability could be subsumed by the logical aspects as provided by the topos approach. 

The non-quantum aspects of probability, which allow for a lack-of-knowledge interpretation, are encoded mathematically by mixed states. These are (finite or infinite) convex combinations of vector states which give density matrices resp. normal states. (Non-normal states like KMS states would require another treatment.) This secondary level of probability can be handled in a manner very similar to ordinary probability theory.

In this way, the topos approach can potentially lead to a clear distinction between the quantum aspects of probability, which may be deducible from the logical aspects, and the non-quantum aspects, which are basically standard probability theory.

We remark that at least for the case $\N=\BH$, it is straightforward to reconstruct the vector $\psi$ from the pseudo-state $\wpsi$ up to a phase, and from $\psi$ we can define the measure $\mu_\psi$, of course. In this sense, the probabilistic part of quantum theory actually can be seen as secondary and derived from the logical, topos-internal aspect. In the future, we will seek for a more direct relation between $\wpsi$ and $\mu_\psi$, ideally in a topos-internal way.


\subsection{Expectation values from measures}			\label{SubsecMeasuresExpVals}
We briefly sketch the connection between measures on clopen subobjects and quantum-theoretical expectation values. Many aspects of this theory need further development and will be treated in detail in a future publication.

Consider a proposition ``$A\epsilon\Delta$'' about the value of the physical quantity $A$ of a quantum system. The proposition corresponds to a projection $\P=\hat E[A\epsilon\Delta]$ via the spectral theorem, and this gives a clopen subobject $\ps{\delta(\P)}$ of $\Sig$ by definition (\ref{DefDasOfProjs}).

Let $\rho$ be a state of the quantum system. In ordinary quantum theory, one assigns a probability of being true to the proposition ``$A\epsilon\Delta$'', given by the expectation value of the projection $\P$:
\begin{equation}
			\mathcal{E}(\P;\rho)=\rho(\P)\in [0,1].
\end{equation}
If $\rho(\P)=0$, then the proposition represented by $\P$ is regarded as (totally) false in the state $\rho$, if $\rho(\P)=1$, then the proposition is (totally) true. In general, for $0<\rho(\P)<1$, the proposition is neither totally true nor totally false.

Let $\mu_\rho$ be the measure on clopen subobjects given by $\rho$. Then
\begin{equation}
			\mu_\rho(\ps{\delta(\P)}):\VN\longrightarrow [0,1],
\end{equation}
and by the definitions of $\ps{\delta(\P)}$ (\ref{DefDasOfProjs}) and $\mu_\rho$ (\ref{Defmu_rho}), this order-reversing function has the value $\rho(\P)$ at all contexts $V$ that contain the projection $\P$. For these contexts, $\dastoo{V}{P}=\P$ and hence $\mu_\rho(\ps{\delta(\P)})(V)=\rho(\P)$, for all other contexts, $\dastoo{V}{P}>\P$ and hence $\mu_\rho(\ps{\delta(\P)})(V)\geq\rho(\P)$. The expectation value $\mathcal{E}(\P;\rho)$ thus is the \emph{minimum} of the function $\mu_\rho(\ps{\delta(\P)}):\VN\rightarrow [0,1]$.

Now consider a self-adjoint operator in the algebra $\N$ of physical quantities of the form
\begin{equation}
			\A=\sum_{i=1}^n a_i\P_i
\end{equation}
for pairwise orthogonal projections $\P_i\in\PN$. The expectation value of $\A$ in the state $\rho$ is given as
\begin{equation}
			\mathcal{E}(\A;\rho)=\rho(\A)=\rho(\sum_{i=1}^n a_i\P_i)=\sum_{i=1}^n a_i\rho(\P_i).
\end{equation}
To each projection $\P_i$, there correspond a clopen subobject $\ps{\delta(\P_i)}$ and an order-reversing function $\mu_\rho(\ps{\delta(\P_i)}):\VN\rightarrow [0,1]$ such that, according to the arguments above, $\rho(\P_i)=\operatorname{min}_{V\in\VN}\mu_\rho(\ps{\delta(\P_i)})(V)$. We obtain
\begin{equation}			\label{ExpVal}
			\mathcal{E}(\A;\rho)=\sum_{i=1}^n a_i\underset{V\in\VN}\min\mu_\rho(\ps{\delta(\P_i)})(V).
\end{equation}
This formula expresses the expectation value in a state $\rho$ of a self-adjoint operator $\A=\sum_{i=1}^n a_i\P_i$ in terms of the clopen subobjects $\ps{\delta(\P_i)}$ and the measure $\mu_\rho$. It should be pointed out that every context $V$ that contains $\A$ also contains all the projections $\P_i$. In such a context, the minima of all the functions $\mu_\rho(\ps{\delta(\P_i)}):\VN\rightarrow [0,1]$ are obtained. Using the fact that the context $V_\A=\{\A,\hat 1\}''$ contains $\A$, we can write
\begin{equation}
			\mathcal{E}(\A;\rho)=\sum_{i=1}^n a_i\mu_\rho(\ps{\delta(\P_i)})(V_\A).
\end{equation}
By just considering the minima of the functions $\mu_\rho(\ps{\delta(\P_i)})$, which give us the expectation values, we throw away a lot of information. It is an interesting task for the future to consider the physical content of the additional information contained in the functions $\mu_\rho(\ps{\delta(\P_i)})$.

Finally, let $\hat B\in\N$ be an arbitrary self-adjoint operator. Each self-adjoint operator in a von Neumann algebra can be approximated in norm by operators of the form $\hat B_j=\sum_{i=1}^j b_i\P_i$, that is, \emph{finite} linear combinations of pairwise orthogonal projections. Let $(\hat B_j)_{j\in J}$ be a family of such operators (which all lie in the algebra $\N$) that approximate $\hat B$ such that for every $\epsilon>0$, there is some $j\in J$ with
\begin{equation}
			||\hat B-\hat B_j||<\epsilon.
\end{equation}
One has
\begin{equation}
			||\hat B-\hat B_j||=\max\{||b||\mid b\in\operatorname{sp}(\hat B-\hat B_j)\}.
\end{equation}
For every state $\rho$, we have
\begin{equation}
			-\epsilon<-||\hat B-\hat B_j||\leq\rho(\hat B-\hat B_j)\leq||\hat B-\hat B_j||<\epsilon.
\end{equation}
This implies
\begin{equation}
			-\epsilon<\rho(\hat B)-\rho(\hat B_j)<\epsilon,
\end{equation}
so the expectation values of the operators $(\hat B_j)_{j\in J}$ approximate the expectation value of $\hat B$ in any given state $\rho$. We saw in (\ref{ExpVal}) how the expectation value of finite linear combinations of projections can be expressed using measures on clopen subobjects.

A future task will be the development of an integration theory based upon measures on clopen subobjects. An integral will have to be defined on arrows from the spectral presheaf $\Sig$ to a presheaf of `values' like $\Rlr$. In the topos approach to quantum theory, certain such arrows represent physical quantities, as described in detail in refs. \cite{DI(3),DI08}.


\section{Discussion and outlook}			\label{SecDiscussion}
It is remarkable that the state space of a von Neumann algebra $\N$ (with no summand of type $I_2$) can be described entirely in terms of measures on the clopen subobjects of the spectral presheaf $\Sig$ belonging to the algebra.

On the one hand, this strengthens further the conception of the spectral presheaf $\Sig$ as an analogue of the state space of a classical system. Probability measures on this `space' correspond exactly to the states of the quantum system. The clopen subobjects $\Subcl{\Sig}$ play the role of measurable sub`sets'.

On the other hand, the theory of (states of) von Neumann algebras is often conceived as non-commutative measure theory. \cite{Con94} Potentially, the construction of measures on the spectral presheaf from the states of a von Neumann algebra can add a new point of view to this theory. The focus is shifted from projections in the algebra to subobjects of $\Sig$.

It is surprising how little non-commutativity is needed in our constructions: the spectral presheaf $\Sig$ consists of the Gel'fand spectra of all the commutative von Neumann subalgebras of $\N$, with a restriction function $\Sig_V\rightarrow\Sig_{V'},\;\l\mapsto\l|_{V'}$, whenever $V'\subseteq V$. The definition of a measure $\mu$ on the clopen subobjects (see (\ref{Defmu})) is local in the sense that for all $\ps S\in\Subcl{\Sig}$, the value $\mu(\ps S)(V)=\mu(\ps S_V)\in [0,1]$ only depends on $\ps S_V$. We saw that $\mu(\ps S)(V)=\mu(\ps{\tilde S})(\tilde V)$ holds whenever $\ps S_V\subseteq\Sig_V$ and $\ps{\tilde S}_{\tilde V}\subseteq\Sig_{\tilde V}$ correspond to the same projection $\P$. While $\ps S_V$ and $\ps{\tilde S}_{\tilde V}$ are subsets of different spaces and hence cannot be compared, the projection $\P$ lies in $\PV$ and $\mathcal{P}(\tilde V)$ (and $\PN$, of course) and hence can be compared across contexts. Interestingly, in the proof of the equation $\mu(\ps S)(V)=\mu(\ps{\tilde S})(\tilde V)$ we do not have to refer to $\P$ explicitly. Properties (1.) and (2.) of a measure are local, since both refer to constructions (meet and join of clopen subobjects, addition of order-reversing functions) which are defined stage by stage. The necessary `non-local' relation between the sets $\ps S_V$ and $\ps{\tilde S}_{\tilde V}$ is provided by the fact that a measure $\mu$ maps clopen subobjects to order-reversing functions from $\VN$ to $[0,1]$.  This gives a mild form of `non-locality', which is enough to define a unique finitely additive probability measure $m$ from a measure $\mu$ on clopen subobjects. Non-commutativity does not enter in any direct way. Rather, the relations between the contexts, i.e., the commutative subalgebras of the non-commutative algebra $\N$ of physical quantities, are central to the proof up to this point. Finally, Gleason's theorem for von Neumann algebras, which is a very strong result, can be used to construct a unique state $\rho_\mu$ from $m$.

This shows that from the spectral presheaf and measures on its clopen subobjects, a strong invariant of the von Neumann algebra $\N$ of the system can be reconstructed, namely the whole space $\mathcal{S}(\N)$ of states of $\N$. As mentioned in the introduction, an ambitious aim of the topos approach is to provide a whole new mathematical setting for quantum theory (and possibly beyond), with little or no direct reference to Hilbert space, operators, states etc. Of course, the task is to show that the new formalism captures the full content of quantum theory. Reconstruction results like the one presented here indicate that this may well be possible.

Many interesting open questions remain. Some of them were already discussed in section \ref{SecMeasExpVals}. On the measure-theoretical side, many developments are conceivable. The spectral presheaf $\Sig$ can be seen as a topological space in the topos $\SetVNop$, closely related to the Gel'fand spectrum of an internal commutative $C^*$- or von Neumann algebra, which is an internal locale in a topos of \emph{covariant} functors\cite{BanMul06,HLS08}. This will be explained in detail in a forthcoming paper. The interplay between open subobjects, given by the topology, clopen subobjects and possibly more general measurable subobjects remains to explored. Further points of interest are inner and outer measures as well as the development of an integration theory. The results in ref.\cite{CS08} will be very useful in this.

On the physical side of things, time evolution of states needs to be considered. It would also be useful to characterise measures coming from pure states of the von Neumann algebra $\N$ of physical quantities (which need not be vector states in general). A major task will be to consider measures for composite systems. In which way does entanglement show up? How do measures on the product of the spectral presheaves of two systems look like? In classical probability theory, not every measure on the product of two measurable spaces is a product measure. If there is an analogous phenomenon for measures on the product of spectral presheaves, can these measures be interpreted as coming from entangled states? How much of entanglement can be described in this way, and what is missing?\footnote{Thanks are due to Steve Vickers, who brought up these questions in a recent discussion.} Finally, how do measures corresponding to fermionic and bosonic states look like? We hope to come back to all these questions in future publications.

\textbf{Acknowledgements.} I would like to thank Chris Isham for many discussions and generous support. Many thanks go to Martin Bojowald for inviting me to contribute to this special issue and also for his patience. Discussions with Pedro Resende, Chris Mulvey and Steve Vickers are gratefully acknowledged.

\end{document}